# Signature of magnetic-dependent gapless odd frequency states at superconductor/ferromagnet interfaces

A. Di Bernardo[1], S. Diesch[2], Y. Gu[1], J. Linder[3], G. Divitini[1], C. Ducati[1], E. Scheer[2], M.G. Blamire[1], J. W. A. Robinson*[1]

## Abstract

The theory of superconductivity developed by Bardeen, Cooper, and Schrieffer (BCS) explains the stabilization of electron pairs into a spin-singlet, even frequency, state by the formation of an energy gap within which the density of states (DoS) is zero. At a superconductor interface with an inhomogeneous ferromagnet, a gapless odd frequency superconducting state is predicted in which the Cooper pairs are in a spin-triplet state. Although indirect evidence for such a state has been obtained, the gap structure and pairing symmetry have not so far been determined. Here we report scanning tunneling spectroscopy of Nb superconducting films proximity coupled to epitaxial Ho. These measurements reveal pronounced changes to the Nb subgap superconducting DoS on driving the Ho through a metamagnetic transition from a helical antiferromagnetic to a homogeneous ferromagnetic state for which a BCS-like gap is recovered. The results prove odd frequency spin-triplet superconductivity at superconductor / inhomogeneous magnet interfaces.

1. Department of Material Science and Metallurgy, University of Cambridge, 27 Charles Babbage Road, Cambridge CB3 0FS, United Kingdom

2. Department of Physics, University of Konstanz, 78457 Konstanz, Germany

3. Department of Physics, Norwegian University of Science and Technology, N-7491 Trondheim, Norway

*Corresponding author: jjr33@cam.ac.uk

## Introduction

In 1961 Ivar Giaever reported[1] the energy gap in the excitation spectrum of the underlying metallic band structure of superconducting Al which directly verified the theory of superconductivity developed by Bardeen, Cooper and Schrieffer[2] (BCS). The superconducting density of states (DoS) is determined by the symmetry of the Cooper pair wave function in time-, orbital-, and spin-space. For BCS superconductors such as Al and Nb, the Cooper pairs are in a spin-singlet state in which their wave function is even with respect to exchange of time coordinates − this is so called even frequency superconductivity.

At a superconductor/ferromagnet interface (S/F), in which the magnetization of the F layer is inhomogeneous, an additional odd frequency pairing symmetry is predicted in which the Cooper pairs are in a spin-triplet state with either zero ($S_z=0$) or a net spin projection[3-4] ($S_z=\pm 1$). For simplicity, we will hereafter refer to these different types of Cooper pairs as "spin-zero" and "spin-one" states. The classification "odd frequency" stems from the fact that the wavefunction of the Cooper pairs is odd with respect to an exchange of the time coordinates.

Indirect evidence for the generation of odd frequency triplet states has emerged from measurements of S/F/S Josephson junctions, which demonstrate a coherence length in F that is much longer than predicted for BCS proximity effects[5-11] or in junctions with ferromagnetic insulators with high spin-polarizations[12]. Odd frequency triplet pairing has also been inferred from critical temperature measurements of S/F/F spin-valves[13-19]. The superconducting excitation spectrum for the odd frequency triplet state should be gapless, meaning the net sub-gap DoS should be enhanced rather than suppressed[20-21] at certain energies (voltages). To date, tunneling conductance measurements to detect odd frequency superconductivity have focused on the F-side of metallic[22-24] and oxide[25-27] S/F interfaces for which a superconducting mini-gap appears in the DoS due to the superconductor proximity effect[28], and which has been predicted to contain both even and odd frequency components and to disappear on a length scale of the ferromagnet singlet pair coherence length $\zeta_{F,Singlet}$ of 1-5 nm. However, discriminating between the odd and even frequency components in the DoS has proven to be controversial and so the results have so far been inconclusive. Eschrig and Löfwander theoretically demonstrated that a spin-active interface also enhances odd frequency pairing correlations on the S-side of an S-F interface[29]. Here the superconducting DoS should be easier to detect because, unlike on the F-side, the DoS is not masked by the normal state background.

To isolate the odd frequency contributions with certainty, it is necessary to amplify them. This can be achieved by introducing a spin-active layer – a magnetic layer with a magnetization direction that is non-parallel to the F layer – at the S-F interface[3]. Such a layer can enhance the pairing amplitude of the spin triplet state. The mechanism behind this effect first involves mixing the singlet pairs due to the spin-dependent phases picked up by electron scattering at the S-F interface, which then results in the emergence of a zero-spin triplet state. The presence of a spin-active layer then rotates the triplet pairs in spin-space and spin-aligned triplet pairs form.

In this article we report scanning tunnelling DOS measurements on superconducting films of Au/Nb proximity coupled to an epitaxial film of Ho. These measurements reveal pronounced changes to the subgap DOS in Nb that sensitively depend on the magnetic phase of Ho; by controlling the magnetic structure of Ho we are able to distinguish between singlet and triplet pair correlations in the superconducting DOS of Nb, since through the magnetic phase of Ho we are able to amplify (suppress) the triplet (singlet) components.

**Tuneable magnetism to modulate sub-gap density of states**

To obtain discriminating evidence for the symmetry of the superconducting state at S/F interfaces and to correlate subgap structure to magnetism, we performed scanning tunneling spectroscopy measurements on epitaxial Nb(20 nm)/Ho(9.5 nm)/Nb(6.5 nm) thin films capped with a 3-nm-thick protective layer of Au (see Fig. 1a and for supporting X-ray diffraction data see Supplementary Figs. 1 and 2). The Au protects the structure from oxidation while the 6.5-nm-thick non-superconducting base layer of Nb provides the correct seed for epitaxial growth[30]. The Nb (20 nm)/Ho (9.5 nm)/Nb (6.5 nm) multilayer had a superconducting transition of ~6.6 K and without Ho of ~8.6 K (see Supplementary Fig. 3).

In single crystal Ho a basal plane helical magnetic phase forms below 133 K[31-32], where the helix turn angle, the angle between magnetic moments in adjacent layers, is 30°. Upon application of a sufficiently large field parallel to the basal plane, a metamagnetic transition to a stable ferromagnetic (F) state takes place in which the moments are aligned in plane[33]. We show below and in Ref. *30* that our Ho thin films behave in a similar way.

In Fig. 1b we have plotted the in-plane magnetization vs field loop $M(H)$ for an epitaxial sample at 10 K. The initial magnetization curve shows a transition to a square F hysteresis loop, which is stable over subsequent field cycles. The fact that the initial magnetization curve goes outside the loop demonstrates a phase transition to a F state. The saturation magnetization ($M_s$) reaches ~2026 emu cm$^{-3}$, which is close to the theoretical value of 10.34 $\mu_B$ per atom[34] and shows that the Ho has near bulk properties.

In Fig. 1c we have plotted the zero-field magnetic phase diagram of our Ho. This was created by the following procedure. First, an in-plane magnetic field (which we call the "set field" or $H_{SET}$) was applied and the corresponding magnetization was measured ($M(H_{SET})$). This field was then switched off; therefore, the remanent magnetization $M_r$ due to $H_{SET}$ can be measured. This procedure was repeated with successively larger values of $H_{SET}$ until saturation was achieved. Below $H_{SET}$= 200-250 mT, $M_r$ and $M(H_{SET})$ barely increase, which implies the helix phase in the bulk (Fig. 1c(i)) is robust, but beyond this field, $M_r$ and $M(H_{SET})$ sharply increase indicating the onset of the metamagnetic transition in which the helix deforms and contains a significant F component (Fig. 1c(ii)). Beyond 500 mT, the Ho is F (Fig. 1c(iii)).

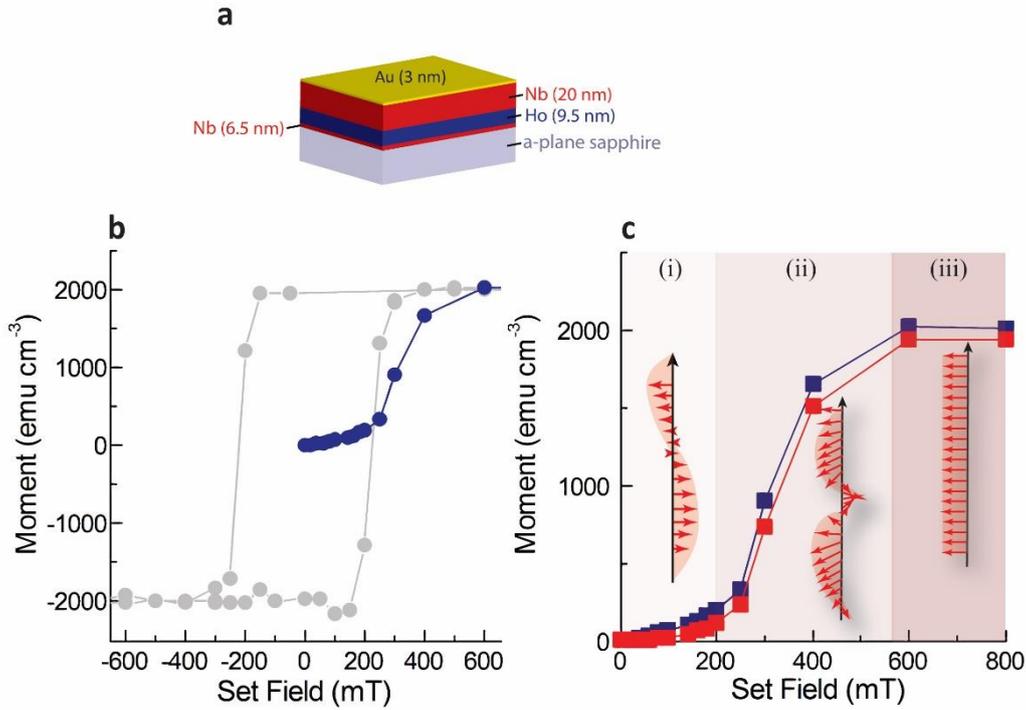

**Figure 1. Magnetic properties of the Au/Ho/Nb multilayer thin films. a,** Illustration of the sample structure. **b**, Initial magnetization curve (blue) and *M(H)* loop (grey) at 10 K. **c**, Magnetization at zero field ($M_r$; red) and with the set field $H_{SET}$ switched on (blue). Lines are guides to the eye. Shaded regions (i-iii) indicate the different magnetic phases of Ho: (i) bulk helix, (ii) coexisting helix and F component, and (iii) F state.

Scanning tunneling spectroscopy was performed at 290 mK with an IrPt tip, mounted on a $^3$He cryostat in which in-plane magnetic fields of up to 500 mT could be applied [35]. The spatial distribution of the superconducting-related spectral features was investigated and, in addition to BCS gaps, two different subgap structures consistently appeared in the DoS in the as-cooled tunneling conductance *dI*/*dV* vs bias voltage *V* as shown in Figs. 2 and 3; *dI*/*dV* is normalized to the normal state conductance at high bias (5 mV). The most common non-BCS subgap structure consisted of a double peak in *dI*/*dV* at symmetrical values of bias voltage (which we name the "double peak" spectra; see Figs. 2a and 3e). In other areas a conductance peak at zero bias voltage was obtained (which we name the "zero peak" spectra; see Figs. 2b and 3f); however, this feature was far less common (in about 10% of scans).

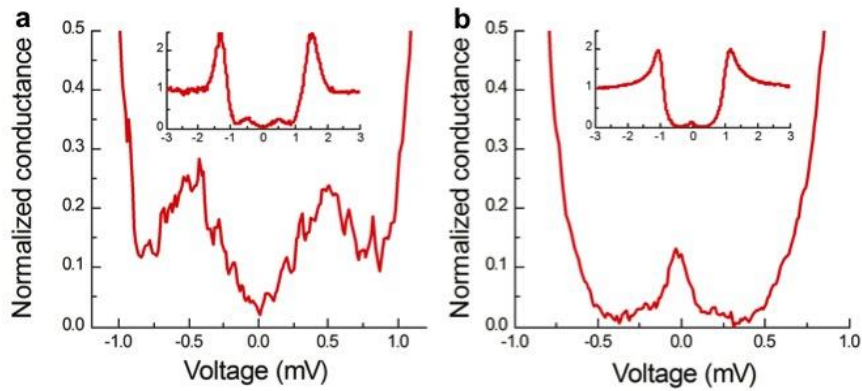

**Figure 2. Normalized zero-field subgap structure in the tunneling conductance *dI/dV* vs bias voltage *V* at 290 mK. a**, double peak and **b,** zero peak enhancements. Insets show the full spectra.

## Subgap dependence on film surface and magnetic phase

To rule out the possibility of spurious effects from surface defects as an explanation of the subgap structure and to ensure the subgap structure is related to the superconductor proximity effect, we measured the spatial dependence of the tunneling DoS and searched for correlations between the spectral features and the surface topography (Fig. 3): as shown, the subgap structure is sensitive to the surface topography and no subgap structure was obtained in regions with a high concentration of defects, showing up in the topography map as increased noise or protrusions and as abundance of black areas in the spectral map (see, for example, around $X = $ -20 nm and $Y = $ -40 nm in Fig. 3a and Fig. 3b). Around grain boundary-like regions we observed either BCS-like gaps (often shallow) or metallic structure, but on defects we only observed noisy metallic spectra. For further details on how the spectral map in Fig. 3b is obtained from the measured DoS spectra, see Supplementary Fig. 4 and Supplementary Note 1.

Conductance maps at different voltages (0 mV, -0.35 mV, -0.5 mV and -1.5 mV) were also determined from the 16x16 matrix of DoS spectra recorded on the sample area shown in Figure 3 (see Supplementary Figs 5 and 6). The maps show a good reproducibility of the local spectral features (Figs. 3c to 3f), which is in agreement with the spectral map reported in Fig. 3b.

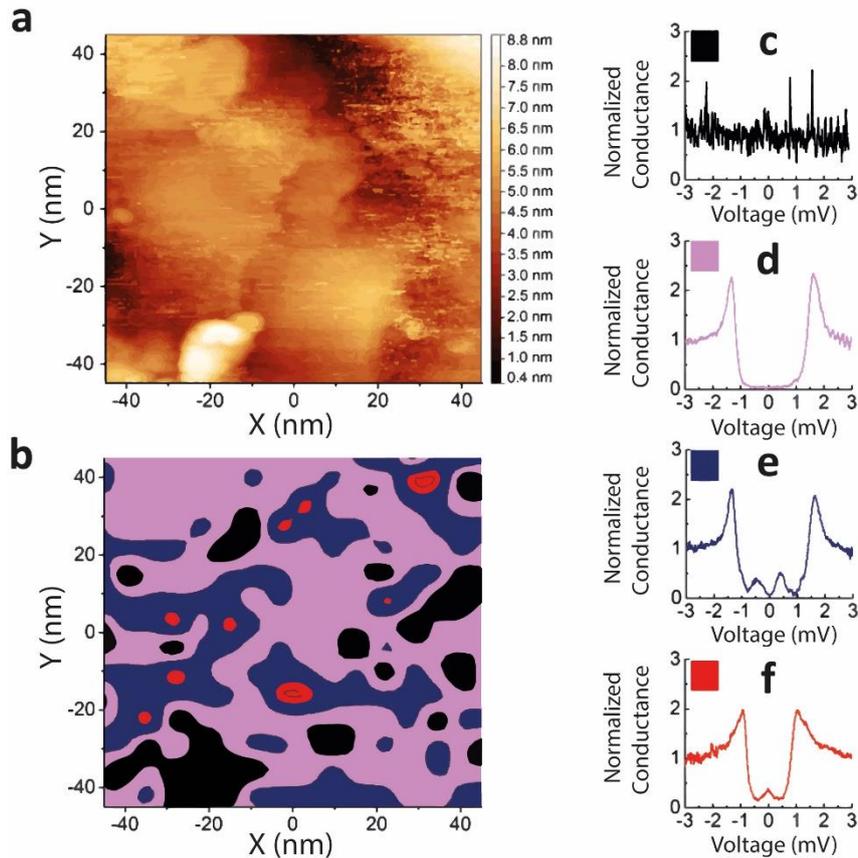

**Figure 3. Spatial dependence of the sub-gap density of states. a**, Surface topography of the Au(3 nm)/Nb(20 nm)/Ho(9.5 nm)/Nb(6.5 nm)/sapphire samples and in **b**, a spatial map of the corresponding spectra obtained at 290 mK; the different colors correspond to the different types of spectra observed and match the spectra shown in **c-f**.

We attribute the BCS-like spectra to positions where the Ho magnetic structure is disturbed such that no pronounced non-collinearity of the magnetization is present. While we cannot confirm this, to be certain that the BCS gap structure is not related to discontinuities of Ho and therefore to the absence of a magnetic proximity effect, we performed transmission electron microscopy (TEM) on cross sectional lamellae of the specimen prepared by focused ion beam (FIB): see Fig. 4 and Methods for further details. Low-magnification scanning transmission electron microscopy (STEM) images (Fig. 4a,b) show that the Ho layer is continuous, which is further supported by energy-dispersed X-ray data (see Fig. 4c-e). High-resolution TEM (Fig. 4f-i) shows that the Nb and Ho layers are crystalline, with little evidence of interdiffusion. Note that during the fabrication of the TEM lamella the top half of the Nb layer and the entire Au layer are damaged and amorphized due to a combination of $Ga^+$ ion implantation occurring during the FIB processing and the deposition of a Pt/C capping layer.

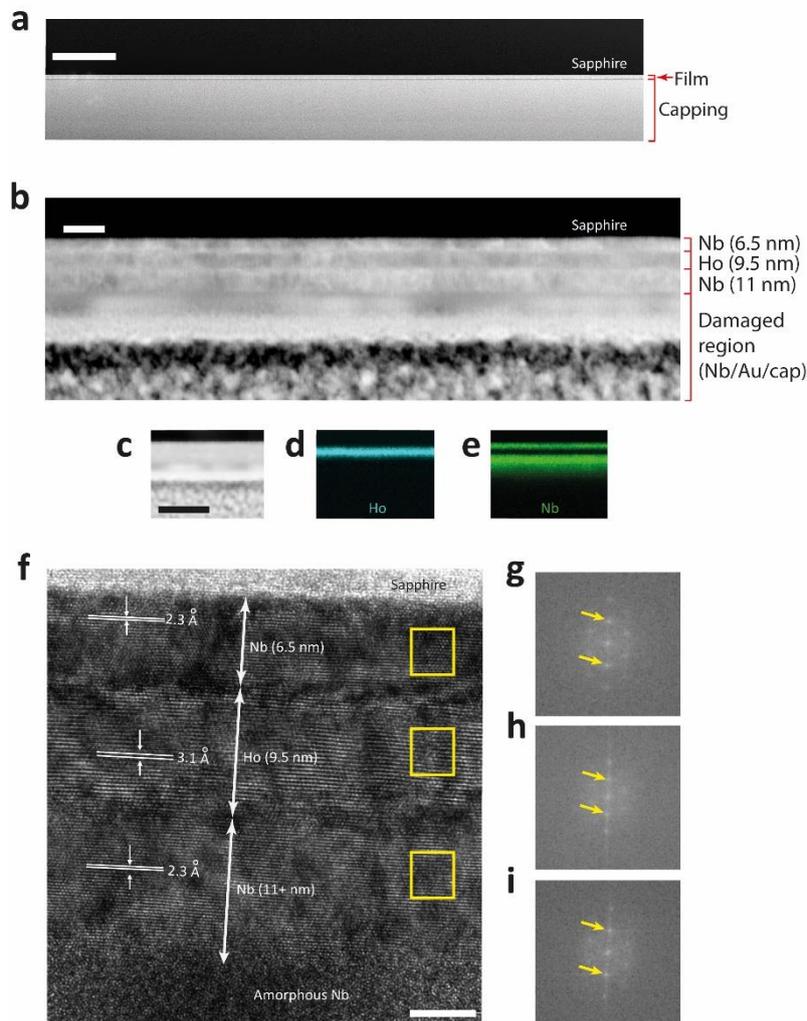

**Figure 4. Electron microscopy of Au/Nb/Ho/Nb lamellas. a** and **b**, Scanning transmission electron microscopy (STEM) images showing continuity of the films. **c**, A STEM image with corresponding energy-dispersed X-ray (EDX) maps highlighting Ho (**d**, blue) and Nb (**e**, green) layers. **f**, A high-resolution transmission electron micrograph of the lamella and corresponding reciprocal space information (fast Fourier transforms) measured in Nb (**g** and **i**) and Ho (**h**). In **g** and **i** the arrows indicate reflections from (110) planes, while in **h** arrows indicate (002) reflections.

The metallic spectra (Fig. 3c) might indicate surface contamination that suppresses superconductivity locally or that disables proper phase locking of the lock-in measurement. Importantly, the subgap structures (Fig. 2) do not appear preferentially at defects, but are more pronounced in clean regions, which is consistent with the presence of a stronger superconductor proximity effect.

Finally, we also tested control samples of Au (3nm) on epitaxial Nb (20nm) without Ho. On these samples only BCS gaps or metallic structure were ever observed (see Supplementary Fig. 7), which confirms that the subgap structure observed in Figs 2 and 3 are related to an unconventional proximity effect due to Ho interacting with Nb.

The most striking evidence that the subgap structure is related to the interaction of Ho with Nb and therefore to an exotic proximity effect was obtained by monitoring the dependence of the subgap features to the magnetic structure of Ho by applying magnetic fields. We first focus on the effect that magnetic fields have on the double peak spectra (similar to that shown in Figs. 2a and 3e). In Fig. 5 we have plotted such spectra in zero field and following the application of successively larger in-plane set fields $H_{SET}$ which cover the metamagnetic transition of Ho (other fields were also measured but are not shown for clarity). The same field sequence as that used to acquire the magnetic phase diagram in Fig. 1c was applied: (1) spectra were measured with the in plane field on ($H_{SET}$); (2) the field was switched off and spectra in zero field was obtained; steps 1 and 2 were repeated with successively larger values of $H_{SET}$. Spectra with the field on or off show qualitatively similar behavior: a disappearing sub-gap structure in the F phase (500 mT) (Fig. 5f); a field-dependent sub-gap structure below 200 mT where the bulk helix hardly changes with applied field (see Fig. 1c); and the appearance of a zero peak near 150mT.

This behavior is clearest in Fig. 5g where we have plotted the difference ($\Delta G$) in the differential conductance obtained at double peak voltage values to the differential conductance obtained at zero voltage as a function of $H_{SET}$. Positive values of $\Delta G$ indicate dominant double peaks while negative values indicate a dominant zero peak. The trend of $\Delta G$ ($H_{SET}$) can be correlated to the magnetic structure of Ho. For fields between 200-500 mT, the entire magnetic helix in Ho (including surface states) irreversibly deforms and so the degree of magnetic inhomogeneity must decrease causing rapid suppression of the odd frequency components[3,4], meaning that the DoS should resemble a BCS gap, consistent with the spectra shown in Fig. 5f. The spectra between 0 and 250 mT are perhaps the most interesting since the subgap structure shows an inversion in the sign of $\Delta G$, despite the fact that the bulk helix in Ho is largely unaffected by magnetic field (Fig. 1c).

In the low field regime (<250 mT), the helix remains intact at least in the bulk of the Ho (Fig. 1c); however, at the surface of Ho the interlayer antiferromagnetic nearest-neighbor exchange coupling energy abruptly decreases, and the surface stabilizes ferromagnetically (in-plane) over several atomic planes[36,37]. From the $M_r/M_s$ ratio of ~0.08 at 250 mT, we estimate a F region of ~0.5 nm per surface or ~2 atomic planes (the c axis parameter at low temperature is 0.564 nm). Such a surface is characteristic of a spin-active interface and indeed in this field regime a zero-energy peak appears in the DoS which is the sought after signature of the odd frequency spin-one triplet component[20,21].

We stress here that spectra similar to that shown in Fig. 3c,d were unaffected by magnetic field which is consistent with the view that in those region with pure BCS gaps or metallic structure the proximity coupling between Ho and Nb is weak. We performed additional tests to rule out artefacts due to magnetostrictive drifts. For instance, if a double-peak structure vanished by increasing the magnetic field, we scanned the surrounding area to check whether it appeared in a position nearby but none were observed. The double-peak subgap structure of Fig. 5 has been recorded in an area where no zero-bias peak is observed in the as-cooled DoS.

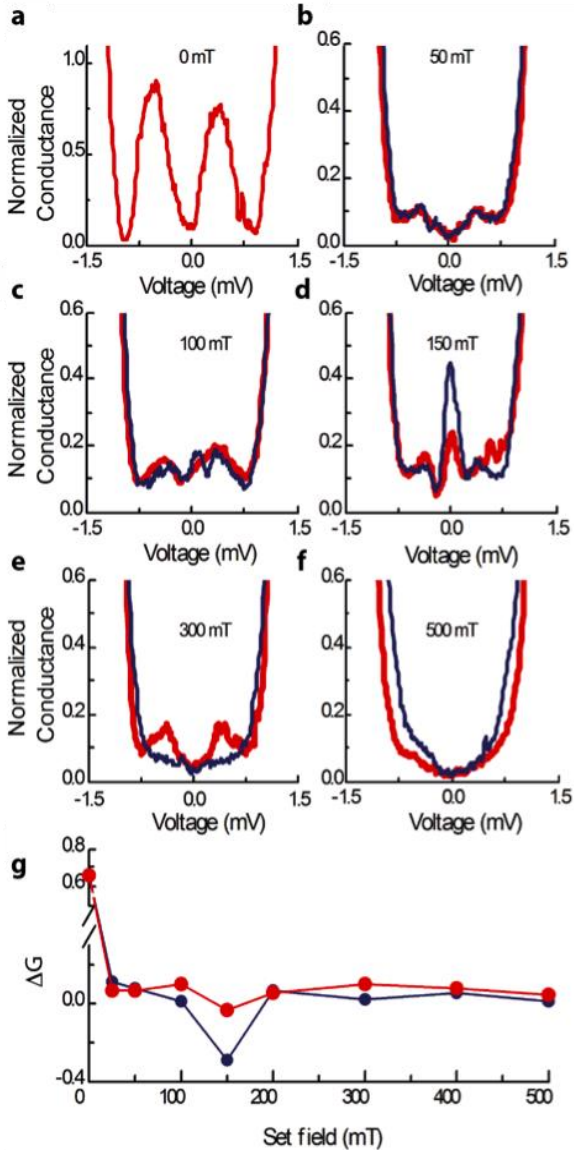

**Figure 5. Effect of magnetic field on the double peak sub-gap density of states. a**, Double peak sub-gap structure in zero field at 290 mK. **b-f**, The effect of switching on (blue) and off (red) successively larger in-plane set fields (as labelled) across the metamagnetic transition of Ho. Data was acquired at the same location. **g**, The difference ($\Delta G$) in the differential conductance obtained at side peak voltages to the differential conductance obtained at zero voltage vs the applied set field $H_{SET}$ for the field off (red; open circles) and on (blue; solid circle).

To understand further the behaviour of zero-energy peaks, we investigated the effect of a magnetic field where zero peaks dominate the as-cooled DoS. As shown in Fig. 6, the zero peak first increases with field before disappearing before the onset of the metamagnetic transition. This is consistent with our hypothesis that such zero-energy peaks are related to the F surface of Ho. This may also explain why zero peaks do not always appear in the as-cooled state as the magnetic noncolinearity at the surface of Ho is likely to spatially vary.

The fact that the zero-energy peak in Fig. 6 first increases with field before decreasing is somewhat consistent with the zero-peak behaviour observed in Fig. 5 (although the initial surface magnetic state must be different in Fig. 5 than in Fig. 6 as the zero field peak was absent in as-cooled state). The behaviour of the zero-peaks in both Figs. 5 and 6 suggests that the degree of non-collinearity between the surface F component and the helix initially increases with field until the field is sufficient to align the surface spins to the top spins of the helix and initiate the metamagnetic transition (> 150 mT).

We also investigated the temperature dependence of a zero peak and the effect of out-of-plane magnetic fields to rule out Kondo as an explanation of their appearance. The zero peaks typically disappear below 1 K (Supplementary Fig. 8), although at higher temperatures the differential conductance becomes noisy since the tip-sample position is unstable. With an out-of-plane field, the magnitude of a zero peak is always found to decrease and none were observed to split with field (Supplementary Fig. 9), which is inconsistent with a Kondo effect. The zero peaks disappear below an out-of-plane field of 500 mT (Supplementary Fig. 9).

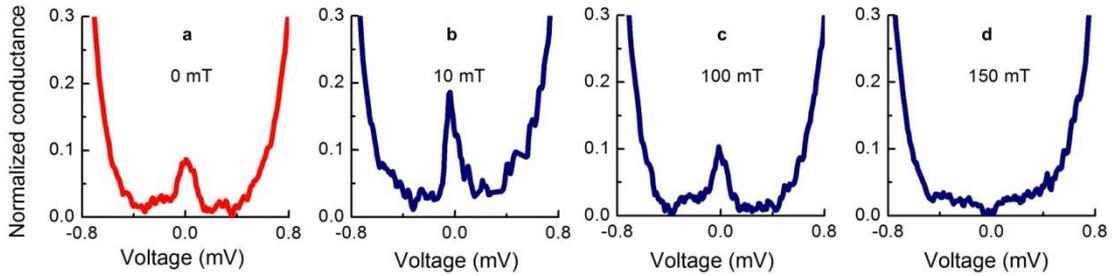

**Figure 6. Effect of magnetic on the single peak subgap structure. a-d**, Zero-peak subgap structure in successively larger in-plane set fields (labelled) at 290 mK. Initially, the amplitude of the zero peak increases with applied field (**a** and **b**) but then rapidly decreases to zero (150 mT) before the onset of the metamagnetic transition of Ho.

## Discussion

The existence of the subgap structure alone indicates unconventional superconductivity; however, to understand more fully the subgap structure and its relation to odd frequency correlations and the zero-energy peak to surface magnetism, we have numerically calculated the DoS using the quasiclassical theory of superconductivity in the diffusive limit and solved in a fully self-consistent way (see Supplementary Note 2). This includes both the suppression of superconductivity near the interface region and the induced odd-frequency pairing in the superconducting region.

Figure 7 shows the obtained DoS in the superconducting region near the Au/Nb interface using the following realistic materials parameters: $R_B/R_S=2$, $R_S = R_F$, $\xi_S=15$ nm, $h/\Delta=3$, $\lambda=3.4$ nm, $\tau_{F,\varphi}=1.3$, $\tau_{S,\varphi}=0.4$, $d_S=20$ nm, $d_F=9.5$ nm, $D_F/D_s = 0.65$, where $R_B$ is the resistance of the barrier between Nb and Ho, $R_{S(F)}$ is the bulk resistance of the Nb (Ho) region, $D_{F(S)}$ is the diffusion coefficient of the Ho (Nb) layer, $\xi_S$ is the superconducting coherence length, $h$ is the exchange field of Ho, $\Delta$ is the superconducting energy gap in Nb, $\lambda$ is the period of the bulk helix in Ho, $d_S$ is the thickness of Nb layer, $d_F$ is the thickness of Ho, and $\tau_{S(F),\varphi}$ describe the spin-dependent interface scattering on the S (F) side of the interface.

In the as-cooled state, we assume that a magnetic helix forms in Ho and we model the effect of surface non-collinearity via spin-active boundary conditions (see Supplementary Note 2). In Fig.7a there is no misalignment between the surface component and the helix and the double-peak structure dominates while in Fig. 7b,c the surface component is misaligned (45° and 90°) with respect to the helix in the basal plane, in which case a dominant zero-energy peak is observed. In the fully F state, the sub-gap structure disappears (Fig. 7d).

The experimental features are qualitatively well described by this model, as non-collinearity between the surface F component and the helix dictates whether the double peak or zero-energy peak structures arise. This model also provides an indirect explanation for why it is more likely to observe the side-peaks than a zero-energy peak in the as-cooled DoS: First, the exact magnetic configuration of Ho is likely to vary across the surface of Ho and, second, that the likelihood of magnetic non-collinearity at the surface is low since ferromagnetic coupling will always prefer a more collinear surface state that favours the double peak structure.

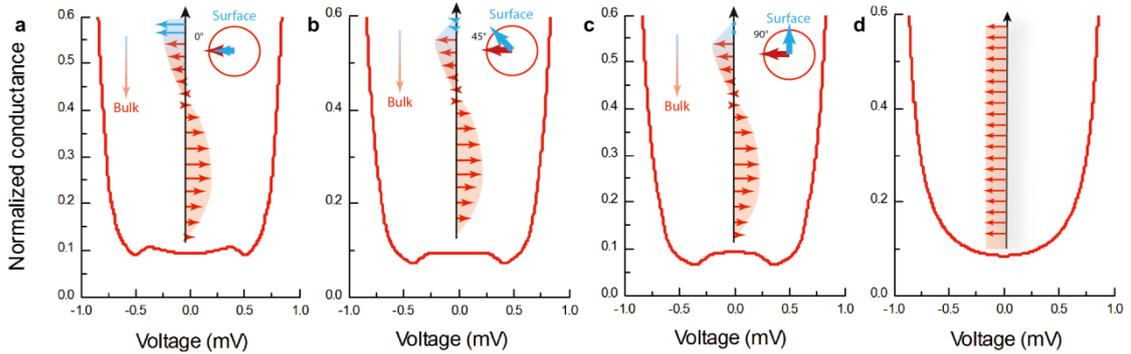

**Figure 7. Calculated magnetic-phase-dependent superconducting DoS in Nb. a**, DoS in Nb for a helical magnetic phase in Ho in which the surface ferromagnetic spins of Ho (indicated by blue arrows) are aligned or **b**, non-collinear by 45° or **c**, by 90° to the bulk helix in Ho (indicated by red arrows). **d**, DoS in Nb for which the Ho has a ferromagnetic phase as sketched showing no subgap structure.

In the model, the sub-gap structure arises due to odd frequency pairing correlations. The crossover from side-peaks to a zero-energy peak occurs in the model as the degree of non-collinearity increases between the surface F component and the helix. This effect could be related to going from dominant spin-zero odd-frequency components in the collinear alignment to a scenario where all triplets, including the spin-one component (relative the interface moment), contribute for the case with strong misalignment. Spin-split peaks inside the subgap region of the superconductor have been shown theoretically to occur in an S/F bilayer when only the spin-zero, odd-frequency component is present[38]. This occurs due to spin-dependent scattering at the interface even in the absence of any magnetic inhomogeneity which induces an effective exchange splitting in the superconductor. On the other hand, the generation of the spin-one odd frequency component is more efficient for the non-collinear alignment between the surface F component and the helix. In this case the zero-energy DoS becomes enhanced due to the gapless nature of odd-frequency pairing and the fact that, unlike the spin-zero odd-frequency component, the electrons in the spin-one pairs are aligned and affected equally by an effective exchange field induced by the spin-active interface so that they should not give rise to any spin-split structure in the DoS. We note that the triplet state observed here is different from that observed in $Sr_2RuO_4$, which is odd parity but even frequency and requires extreme sample purity[39].

In conclusions, we have observed a magnetic-phase-dependent superconducting subgap DoS in NB proximity coupled to Ho. The results demonstrate a profound modification of the superconducting state because of the presence of odd frequency spin-triplet pair correlations. The existence of triplet states in a superconductor is profoundly interesting in its own right. Such available states mean that within the energy gap of a superconductor a finite net spin-polarization exists in the absence of dissipation. One intriguing possibility is that the spin-one states could accommodate a spin accumulation, meaning that in a non-equilibrium S/F device, the S layer could serve as a spin sink or channel to transmit spin in spintronics[40].

## Methods

**Epitaxial thin film growth.** Au/Ho/Nb samples were grown by direct current magnetron sputtering onto heated *a*-plane (110) sapphire substrates. The deposition chamber was continuously cooled via a liquid nitrogen jacket giving a base pressure of approximately $10^{-8}$ Pa with the water partial pressure below $10^{-9}$ Pa as confirmed by an in-situ residual gas analyzer. A base layer of 6.5-nm-thick, non-superconducting Nb (110) was first grown at a substrate temperature of 880 °C using a deposition rate of 0.035 nm s$^{-1}$. This layer was essential to avoid oxidation of Ho and to initiate epitaxial growth. A 9.5-nm-thick layer of epitaxial Ho (002) was then grown on the Nb at 650 °C at a rate of 0.035 nm s$^{-1}$. A 20-nm-thick layer of superconducting Nb (110) was then grown on Ho. Finally, substrates were allowed to cool to room temperature in 1.5 Pa of Ar and finally were capped with a 3-nm-thick layer of polycrystalline Au. While Au protects the underlying structure from oxidation, it also wets well to Nb and forms a highly stable intermetallic compound with Nb (and therefore at the Nb/Au interface in our experiment).

**Scanning tunnelling microscopy.** The density of state (DoS) spectra were acquired using a custom-built scanning tunnelling microscope (STM) in spectroscopy mode with an IrPt tip, mounted on a $^3$He cryostat. The STM consists of one segmented piezo tube for scanning and coarse approach, using slip and stick motion. An in-situ superconducting magnet can apply an in-plane magnetic field of up to 500 mT. The electronic system has a voltage resolution of roughly 20 µV at base temperature (240 mK). In-depth instrument details are described in Ref.

35 and its suitability for studying superconducting proximity effects has been demonstrated in Ref. 41.

The topography maps were recorded by slowly scanning in x direction. The images contain 256 x 256 data points. For recording the differential conductance $dI/dV$ spectra the scan was stopped, and the feedback loop was switched off at a current of 300 pA and a voltage of 3 mV. The spectral map Fig. 4b contains 16 x 16 spectra. The spectra were acquired using a lock-in amplifier. The direct-current-bias-voltage was modulated with an A.C. voltage with root-mean-square amplitude of 14 µV and frequency of 733 Hz The $dI/dV$ was measured using a current-to-voltage converter with a gain of 1 V/nA and the measuring time per spectrum was typically around 1-2 minutes. All spectra were acquired at a tunnelling resistance of 10 MΩ where the signal-to-noise ratio was optimal while avoiding distortions of the spectrum due to the presence of the counter electrode. This tunnelling resistance is the value that we ordinarily adopt to perform STM measurements with our setup.

For higher temperature measurements, the STM was allowed to stabilize for at least 5 minutes at each new temperature before spectra were acquired. All spectra are normalized with respect to the ohmic conductance measured at 5 mV and so well beyond the superconducting gap edge and voltage offsets were subtracted using the symmetry of the spectra. To ensure reproducibility of the observed superconducting-related spectral features and minimize noise contribution, each differential conductance spectrum is the average of three scans performed on the same sample spot.

**Transmission electron microscopy.** Lamellae for TEM analysis were prepared using a FEI Helios Nanolab focused ion beam/scanning electron microscope (FIB/SEM). A 2µm-thick capping layer was deposited to protect the film. STEM-HAADF (High-Angle Annular Dark Field) images and EDX maps were acquired on a FEI Osiris (200 kV acceleration voltage) equipped with a high-brightness field emission gun (X-FEG) and a Bruker Super-X detector.

**Supplementary Information** available at:
http://www.nature.com/article-assets/npg/ncomms/2015/150902/ncomms9053/extref/ncomms9053-s1.pdf

**Acknowledgments** The work was also funded by the Royal Society through a University Research Fellowship held by J.W.A.R., the Leverhulme Trust (J.W.A.R., E.S. and M.G.B.) through an International Network Grant (IN-2013-033),and the EPSRC through NanoDTC EP/G037221/1 (A.D.B and J.W.A.R.). Further support was received from ERC AiG "Superspin." (M.G.B.). Y.G. acknowledges financial support from King's College, Cambridge. A.D.B. acknowledges additional financial support from the Schiff Foundation and Cost Action MP1201. J.L. was supported by the "Outstanding Academic Fellows" programme at NTNU and the Norwegian Research Council grants 205591 and 216700. Useful discussion with K. Halterman and F. S. Bergeret are acknowledged. G. D. and C. D. acknowledge funding from the ERC under grant number 259619 PHOTO-EM. C.D. acknowledges financial support from the EU under grant number 312483 ESTEEM2.

**Author Contribution** J.W.A.R. had the idea of the experiment and supervised the project. A.D.B. did the STM measurements with the help of S.D. under the supervision of E.S. Y.G. grew the epitaxial films and characterized their magnetic properties under the supervision of M.G.B. The theoretical model was written by J.L. Electron microscopy was performed by G.D. under the supervision of C.D. J.W.A.R. and A.D.B. analyzed the measurement data and wrote the paper with contributions from all the other authors.